# Nearly Free Electron States in Graphene Nanoribbon Superlattices

Shuanglin Hu[1], Zhenyu Li[1], Qiaohong Liu[1], Xudong Xiao[2], J. G. Hou[1], and Jinlong Yang[1]

[1]Hefei National Laboratory for Physical Sciences at Microscale, University of Science and Technology of China, Hefei, Anhui 230026, China
[2]Department of Physics, The Chinese University of Hong Kong, Shatin, New Territory, Hong Kong, China

jlyang@ustc.edu.cn

ABSTRACT

Nearly free electron (NFE) state is an important kind of unoccupied state in low dimensional systems. Although it is intensively studied, a clear picture on its physical origin and its response behavior to external perturbations is still not available. Our systematic first-principles study based on graphene nanoribbon superlattices suggests that there are actually two kinds of NFE states, which can be understood by a simple Kronig-Penney potential model. An atom-scattering-free NFE transport channel can be obtained via electron doping, which may be used as a conceptually new field effect transistor.

# 1. Introduction

Nearly free electron (NFE) state [1-7] exists widely in low dimensional materials. As an important state, it can play a key role in doping processes. For example, its occupation is found to be a determining factor for superconductivity in graphite intercalated compounds.[8] At the same time, NFE state has a long transport relaxation time,[4] which is very desirable for electronics applications.

NFE state was originally found in alkali-metal graphite intercalation compounds, and it was thought arising from alkali *s* electrons.[1] Posternak et al.[2] then proved that it is a general interlayer bonding state which is not correlated with the presence of alkali metal atoms. Recently, it was further argued that NFE states are Rydberg-like states bound by image potential.[9] Therefore, despite of intense studies, controversy on the physical origin of NFE state still exists in the literature. We note that previous studies mainly focus on a few low-lying NFE states below the vacuum level. A systematic study on the whole NFE state manifest may help us to understand them better.

NFE states are usually far from the Fermi level. To utilize them, we need to lower the energy of NFE state to the vicinity of the Fermi level. Alkali metal doping is the most commonly used method to occupy NFE states,[10] where it was originally suggested that the band hybridization with the *s* orbital of alkali metal lower the energy of NFE states.[11,12] A recent density functional theory (DFT) calculation revealed that pure electron doping can also lower the NFE states,[3] where an NFE state deformation upon electron doping was suggested.

As a two dimensional material, graphene[13] may replace silicon in the next-generation electronics.[14] For this purpose, it is desirable to open an energy gap in the graphene band structure. Cutting it into thin strips, known as graphene nanoribbons (GNRs), can open a gap in the otherwise gapless system. However, the electronic structure of GNR strongly depends on its orientation, width, and edge structure,[15-17] and their precise control remains a formidable challenge.[18] It is thus very desirable to explore new states in GNR related system which are not sensitive to the detailed atomic configuration of the host material.

In this paper, based on plane-wave first-principles calculations, we identify two distinct types of NFE states in GNR superlattice. They have very different spatial distributions, hence their response to electron doping are dramatically different. One of them is distributed in the vacuum area between GNRs and can be occupied easily upon electron doping. Such a type of occupied NFE state represents an ideal one dimensional electron system (1DES), and maybe has a great potential in electronics applications.

**2. Methods and model**

We performed electronic structure calculations within the density-functional theory (DFT) with the PW91 generalized gradient approximation (GGA).[19] The Vienna *ab initio* simulation package (VASP) [20,21] based on plane-wave basis set was used, with a cut-off energy of 400.0 eV. The projector augmented wave (PAW) [22] method was used to describe the electron-ion interaction. Total energies were converged to $10^{-5}$ eV. The graphene nanoribbon superlattices were modeled by the periodic boundary condition. Edges of all nanoribbons were saturated by H atoms. The vacuum layers between neighboring images along *y* and *z* directions were larger than 20 Å.[see Fig. 1(a)] Special *k*-points for Brillouin zone integrations were sampled using the Monkhorst-Pack scheme[23] with 25 and 19 points along the zigzag and armchair nanoribbon direction, respectively. In *y* and *z* directions only the $\Gamma$ point was used. Electrons and holes were injected by adding or removing electrons from the system and using a jellium background to maintain charge neutrality.

## 3. Results and discussion

First, we use either N=12 zigzag graphene nanoribbon (ZGNR) or N=21 armchair graphene nanoribbon (AGNR) to build two superlattice systems. Both types of ribbons have a width about 26.4 Å. All ribbons are put along the *x* axis in the *x-z* plane. Neighboring GNRs in the *z* direction are separated by a distance about 33.6 Å. Band structures and $\Gamma$ point density profiles of several lowest NFE states for these two systems are presented in Figs. 1(b) and 1(c), respectively. We find that there are two types of NFE states. The first type is denoted as NFE-ribbon states. These states mainly distribute on the two sides of the ribbon plane, and their energies are more than 3.0 eV higher than the Fermi level. As shown in Fig. 1, the NFE-ribbon states have increasing number of nodal planes in the *z* direction with the increase of energy, and a quantum number *n* can be assigned accordingly. In the *y* direction, these NFE-ribbon states show no nodal plane, which means that in this dimension they have the lowest quantum number *m*=0. The second type of NFE states appear at higher energy, they are denoted as NFE-vacuum states. They mainly distribute in the vacuum region between nanoribbons and have energies more than 4.0 eV above the Fermi level, higher than the vacuum level. In higher energy range, increasing numbers of nodal planes in both *z* and *y* directions for NFE-vacuum states can be observed.

As shown in the band structures presented in Figs. 1(b) and 1(c), both types of NFE states have a parabolic dispersion from Γ to X, similar to NFE states in nanotubes.[4] The effective mass at Γ in the ribbon direction is in the range of 1.15 and 1.26 $m_0$ for low energy NFE-ribbon states and around 1.10 $m_0$ for NFE-vacuum states, where $m_0$ is the mass of a free electron. The effective masses obtained here are also close to those in nanotubes.[4, 24]

To study the origins of these NFE states, we examine the electrostatic potential. In the *y* direction, the *x-z* plane averaged electrostatic potential presents a narrow and deep (about 10 eV) well. Such a potential confines the NFE-ribbon states close to the two sides of ribbon plane. In the *z* direction, the *x-y* plane averaged potentials are plotted in Figs. 2(a) and 2(b). They are basically square potential wells if we ignore the oscillations at atom positions. A one-dimensional (1D) Kronig-Penney potential[25] [see Fig. 2(c)] is thus a reasonable model for the periodic potential in the *z* direction. The energies and spatial distributions of states in this model can be solved analytically.

The lower potential $V_1$ in the Kronig-Penney model corresponds to the ribbon region, while $V_2$ represents the vacuum region. Two types of states can be found in this model as shown in Fig. 2(d). The first type of states starting from $V_1$ mainly distribute in the ribbon region. In the energy range [$V_1$, $V_2$], there are several energy bands, corresponding to the increasing number of nodal planes *n*. The energy dispersion of these bands along *z* direction is very small. The second type of states mainly distributed in the vacuum region can be found above $V_2$. Their energies equal to the resonance energies above this potential, which can be related to an infinite square potential well with the bottom of the potential equals to $V_2$,

$$E_n = V_2 + \frac{\hbar^2}{2m}\left(\frac{n\pi}{W_2}\right)^2, \qquad (1)$$

where $W_2 = a-b$ is the width of the vacuum region [see Fig. 2(c)]. The spatial distribution of these two types of states in the 1D Kronig-Penney model resembles the NFE-ribbon and NFE-vacuum states in DFT calculations, respectively.

From the analytical model, the energy spacing for the first type of states decreases as the ribbon width increases. When the ribbon width becomes infinite, the energies of NFE-ribbon states should converge to the energy of NFE states of two-dimensional graphene with the same quantum number *m* in *y* dimension, but the discrete quantum number *n* in the *z* dimension will become a continuous wave vector $k_{//}$. If we increase the width of the vacuum region for a given type of nanoribbons, the energies of the second type of states will downshift to the vacuum level ($V_2$) while the energies of the first type of states stay almost unchanged. All these trends are consistent with the results obtained by our DFT calculations.

The different energy dependences of the two types of NFE states on the widths of ribbon and vacuum region in the *z* direction indicate that they have different origins. The NFE-ribbon states are bound to the surface plane of GNRs and should have the same origin as the NFE states in 2D graphene.[26,27] The NFE-vacuum states are not an intrinsic character of a single GNR, but a result of quantum confinement from the periodic superlattices.

It is interesting to see how these two types of NFE states respond to electron doping. Electron doping can modify the potential and move the NFE states to the Fermi level.[3] Band structures and Γ point density profiles of some NFE states are calculated at various electron doping levels. As shown in Fig. 3, when electrons are doped, NFE states are downshifted significantly in energy, while the localized σ* and π* states undergo little change. With respect to the Fermi level, the NFE-vacuum states are more sensitive to electron doping than NFE-ribbon states and drop more rapidly in energy. Such a difference between these two types of NFE states is due to their different spatial distributions. With electron doping, the whole potential decreases toward the Fermi level. In the light doping case [see Figs. 3(a) and 3(c)], the potential in the vacuum region decreases more rapidly than that in the ribbon region, which leads to the faster dropping of NFE-vacuum states in energy.

When more electrons are doped, as shown in Fig. 3(b), some NFE states start to be occupied. The occupied NFE states are all NFE-vacuum states and the extra electrons thus distribute mainly in the vacuum region. At the doping concentration of 0.04 electrons per carbon atom, the electron gas concentration of the occupied NFE-vacuum states is about $3\times10^{18}$ cm$^{-3}$, similar to the carrier concentration in doped silicon. Although the whole electrostatic potential is further reduced to approach the Fermi level upon heavier doping, compared to the light doping case, the potential in the vacuum region increases relative to the potential at the ribbon region [see Fig. 3(d)]. It is the relative potential increase in the vacuum region that hinders further decreasing of the energy of NFE-vacuum states. This is also the reason why the lowest NFE states varies slowly with electron doping level after being occupied, as shown in Fig. 4(a). We note that in the carbon nanotube case,[3] NFE state deformation was used to explain the different appearances of the lowest NFE states upon electron doping. According to our results, there should also be two types of NFE states in nanotube bundles, and they respond to electron doping differently.

Occupied NFE states can provide an ideal ballistic transport channel. For practical application, it is important to study the edge sensitivity of these NFE states. As shown in Fig. 4(a), before it reaches the Fermi level, the energy of the lowest NFE-vacuum state in the N=21 AGNR superlattice drops slightly faster than that of the N=12 ZGNR superlattice. However, after the electron doping level is larger than 0.02 electrons per carbon atom, the energies of the lowest NFE-vacuum states in AGNR and ZGNR superlattices are similar, and both begin to drop very slowly with the electron doping level. Such edge insensitivity makes the application of the occupied NFE-vacuum states practical.

We also note that, when we fix the vacuum width, the minimum electron doping concentration to move the lowest NFE-vacuum state to Fermi level in GNR superlattice decreases with the increase of ribbon widths [see Fig. 4(b)]. This result encourages to use wide GNRs which are easy to obtain than narrow ribbons in experiment. When the ribbon width is about 9 nm, the critical doping level can be as low as 0.013 electrons per carbon atom, corresponding to a concentration of $4.9 \times 10^{13}$ $cm^{-2}$. A previous Raman spectrum experiment on graphene samples obtained by micromechanical cleavage clearly showed the existence of excess charges with concentrations up to about $10^{13}$ $cm^{-2}$, even in the absence of intentional doping.[28] Therefore, the doping concentration required to occupy NFE-vacuum states is only moderate for superlattices composed of wide GNRs.

To further check the stability of such NFE states and their response to electron doping, we consider boron/nitrogen doped GNR superlattices. The optimized geometries of the doped N=12 zigzag GNRs are shown in Figs. 5(a) and 5(d), respectively. From the band structure and Γ point density profiles [Figs. 5(b) and 5(e)], it is clearly seen that boron/nitrogen doped zigzag GNR superlattice has similar NFE states compared with the corresponding pristine GNR superlattice. With electron doping, we have also observed similar downshifting behavior of the NFE-vacuum states, which can be occupied at some points [see Figs. 5(c) and 5(f)]. Boron/nitrogen doped armchair GNR superlattice gives similar results. Therefore, the electron doping induced occupation of NFE-vacuum states is not affected by slight boron/nitrogen doping, and it is thus a very robust behavior.

Now, it is safe to conclude that the predicted electron doping induced NFE-vacuum state occupation is very practical. Therefore, it is possible to use NFE-vacuum states in GNR superlattices as an ideal ballistic transport channels to construct a graphene based field effect transistor (FET), since NFE-vacuum states can be tuned to be occupied at a feasible electron doping level while hole doping shifts their energies far above the Fermi level. In practice, electron doping can be realized by electrostatic gating. New transport channels are expected to appear at a suitable gate voltage [see Fig. 6(a)]. With its drain electrode localized in a vacuum channel, we obtain a conceptually new FET device [see Fig. 6(b)]. Since it utilizes an ideal transport channel, such a GNR superlattice based FET is expected to give high on-off ratio compared to previous FET based on GNR.[29-31] It also has the advantage that the orientation, width, and edge structure of GNR, which is difficult to control, is not critical to the performance of the GNR superlattice FET. The device design principle described here can be directly applied to carbon nanotube bundles, but graphene sheets should be easier to be charged via external gate voltage.

Occupied NFE-vacuum state also provides an ideal 1DES with minimized scattering. 1DES is known to show correlation effect different from its two- and three-dimensional counterparts, and excitations in 1DES are not quasi-free electrons with an effective mass. 1DES can be described by Luttinger liquid theory.[32-34] Although this theory has been established for a long time, its experimental realizations has only been identified more recently.[35-38] The totally new 1DES proposed in this study will provide more flexibilities to study electron correlation in low dimensions.

We note that the choice of GNR as the material for NFE state FET is no arbitrary. For example, our test calculations show that materials with large band gap such as boron nitride nanoribbons (BNNRs) do not show similar NFE-vacuum state behavior. We take the N=21 armchair BNNR (see Fig. 7) as an example, Similar to boron nitride nanotubes (BNNTs), BNNRs are wide-gap semiconductor, and the bottom of conduction band is an NFE state.[24,39] Upon electron doping, the NFE states at conduction band edge (CBE) are partially occupied. Although we also observed the downshift of NFE-vacuum states, however, different with the case of GNR superlattice as shown in Fig. 3(a), the NFE-vacuum states mix with NFE-ribbon states, and the lowest occupied NFE states is still mainly the original NFE-ribbon state at CBE.

## 4. Conclusions

In conclusion, by first-principles calculations, we have identified two types of NFE states in GNR superlattices. The existence of these two types NFE states can be understood from a simple Kronig-Penney model. One type of NFE states mainly distributed in the vacuum region can be tuned to be occupied via electron doping through a gate voltage. Such states provide an extremely clean 1DES and also lead to a new concept of FET, they are thus of both great technical potential and scientific importance.

**Acknowledgements**

The authors are grateful for supports from NFSC (50721091, 20803071, 20873129, and 20933006), the National Key Basic Research Program (2006CB922004), MOE (NCET-08-0521, FANEDD-2007B23), and CAS (KJCX2-YW-W22). Calculations were performed at the USTC-SCC, the Shanghai Supercomputer Center, and the SCCAS. X. X. also acknowledges the support by the Research Grants Council of

**FIGURE CAPTIONS**

Figure 1 (a) The geometric structure of the N=12 ZGNR superlattice. The band structure and the Γ point density profiles of the five lowest NFE-ribbon states (lower panels) and the two lowest NFE-vacuum states (top panels) in the *y-z* plane normal to the nanoribbon for (b) N=12 ZGNR and (c) N=21 AGNR superlattices, respectively. The red and blue lines in the band structures represent NFE-ribbon and NFE-vaccum states, respectively. The gray dot-lines represent normal σ and π states.

Figure 2 The *x-y* plane averaged electrostatic potential for a unit cell of the (a) N=12 ZGNR and (b) N=21 AGNR superlattices, respectively. The Fermi level is set to zero. (c) The one-dimensional Kronig-Penney potential model. $W_1$=b, $W_2$=a-b. (d) Profiles of the norms of states in the Kronig-Penney potential model with $V_1$=2.8eV, $V_2$=4.0eV, *b*=26Å, and *a*=60Å. The red and blue lines represent two types of states, corresponding to the two types of NFE states, respectively.

Figure 3 The band structures and Γ point density profiles of a few lowest NFE states of the N=12 ZGNR superlattice with electron doping level at (a) 0.01 and (b) 0.04 electrons per carbon atom. The red and blue lines in the band structures represent NFE-ribbon and NFE-vacuum states, respectively. The NFE-ribbon states in the 0.04 electrons per carbon atom case are out of the energy range of bands shown in (b). The *x-y* plane averaged electrostatic potential for a unit cell of the N=12 ZGNR superlattice with electron doping level of (c) 0.01 and (d) 0.04 electrons per carbon atom, respectively. The Fermi level is set to zero.

Figure 4 (a) Energy of the lowest NFE state with respect to the Fermi level as a function of the electron doping concentration for the N=12 ZGNR and N=21 AGNR superlattices. The red and blue lines are used as a guide for eye. (b) The minimum electron doping concentration to move the lowest NFE state to Fermi level in ZGNR superlattices with different ribbon widths.

Figure 5 (a) The geometry of B doped N=12 ZGNR. The band structure and Γ point density profiles of a few lowest NFE states of B doped ZGNR (b) with no electron doping and (c) with electron doping level at 0.04 electrons per atom. (d) The geometry of N doped N=12 ZGNR. The band structure and Γ point density profiles of a few lowest NFE states of N doped ZGNR (e) with no electron doping and (f) with electron doping level at 0.04 electrons per atom. The red and blue lines in the band structures represent NFE-ribbon and NFE-vacuum states, respectively.

Figure 6 (a) Electron doping induced transport through NFE-vacuum states. (b) A schematic FET device model based on NFE-vacuum state.

Figure 7 (a) The band structure and Γ point density profiles of the five lowest NFE-ribbon states (lower panels) and the two lowest NFE-vacuum states (top panels) of N=21 armchair BNNR superlattice with no electron doping. The red and blue lines in the band structure represent the NFE-ribbon and NFE-vacuum states, respectively. (b) The band structure and $\Gamma$ point density profiles of the seven lowest NFE states with electron doping level at 0.01 electrons per atom. The purple lines in the band structure represent NFE states with both NFE-ribbon and NFE-vacuum characteristics.

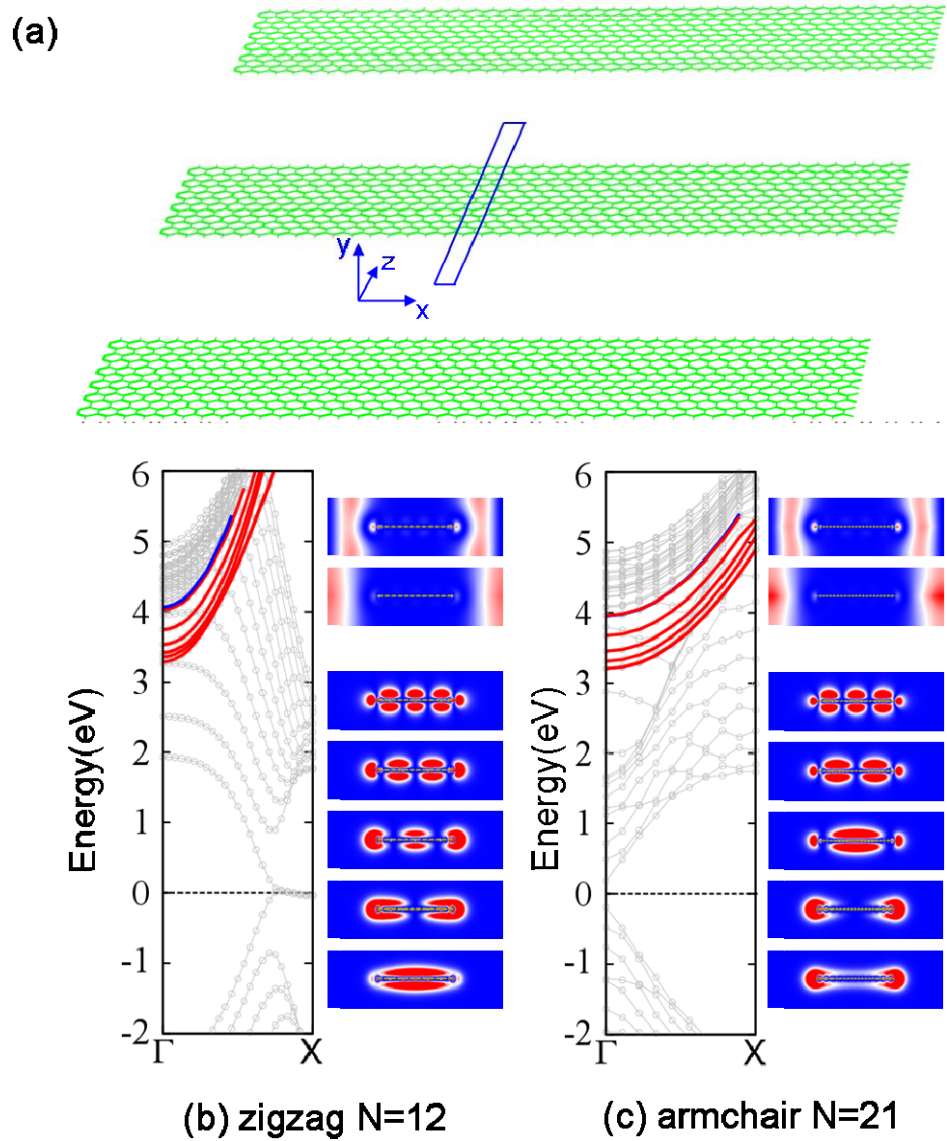

Figure 1

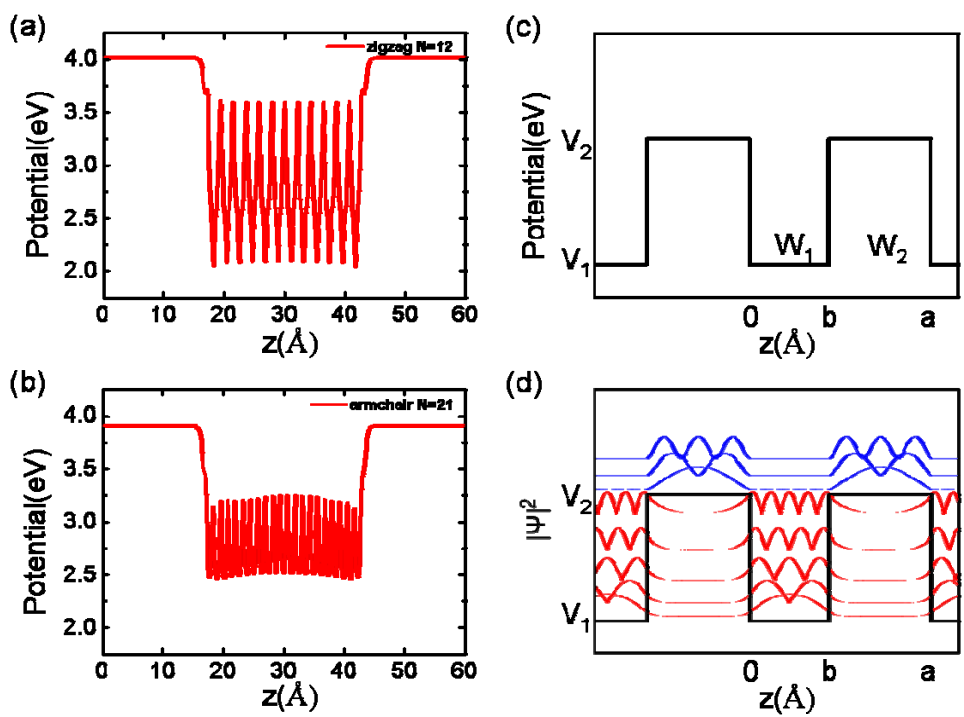

Figure 2

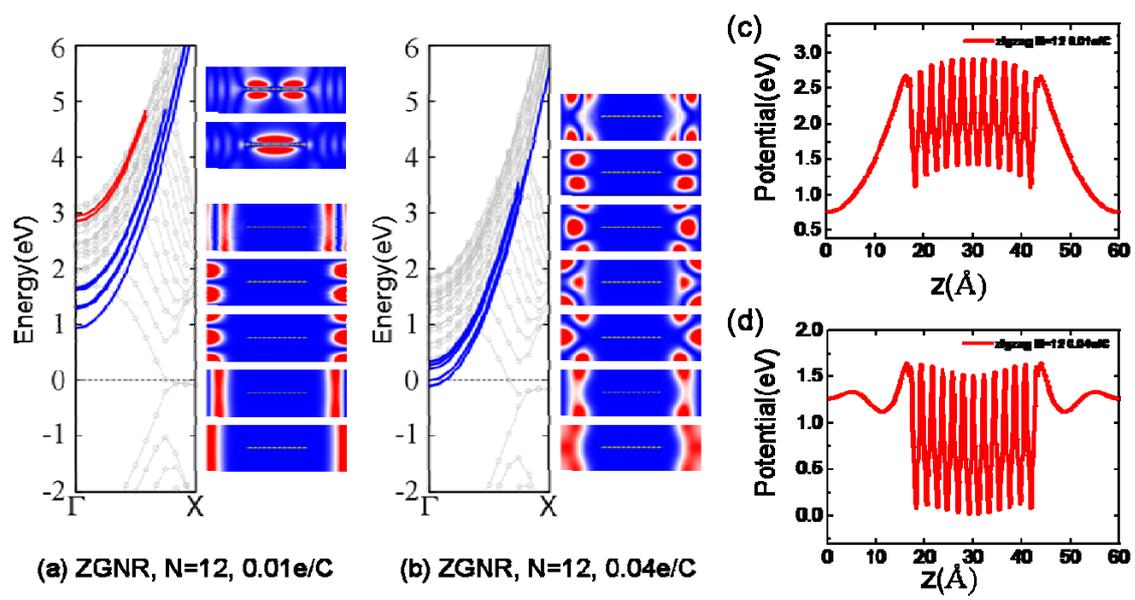

Figure 3

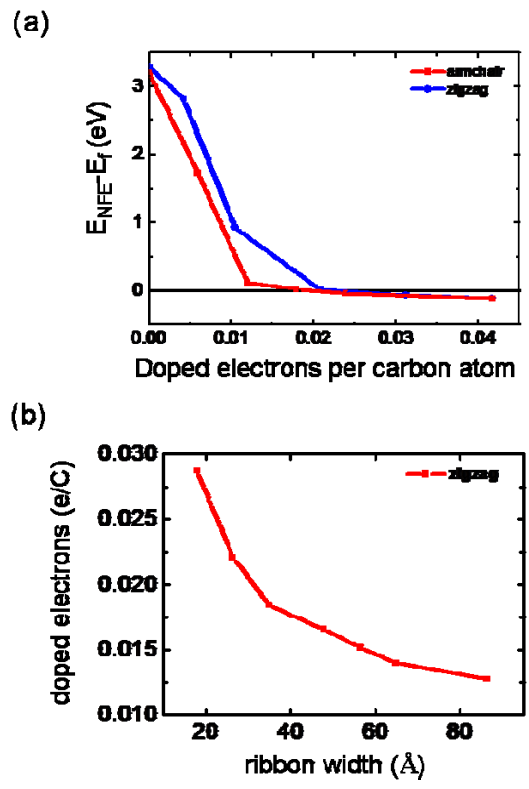

Figure 4

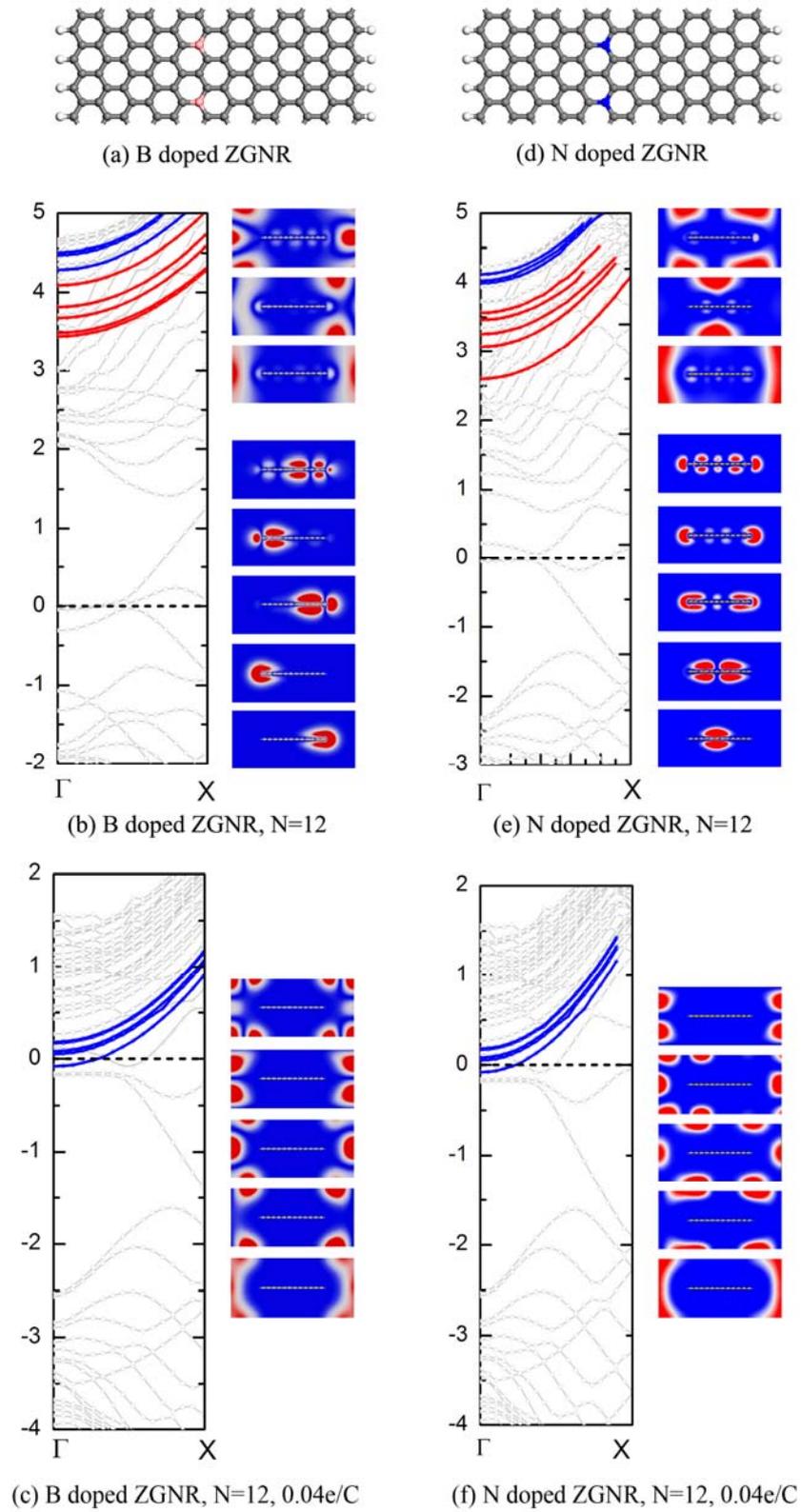

Figure 5

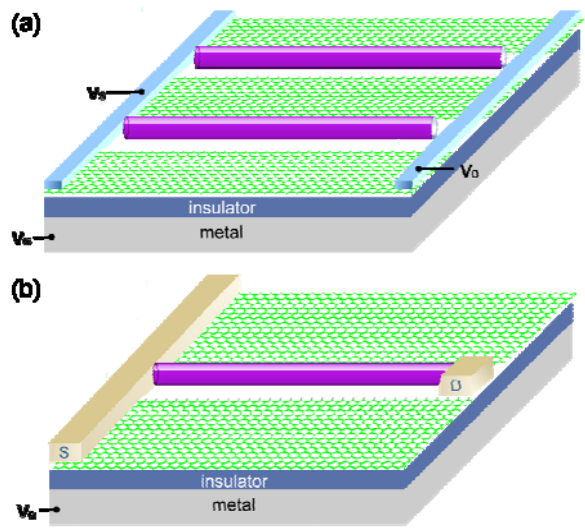

Figure 6

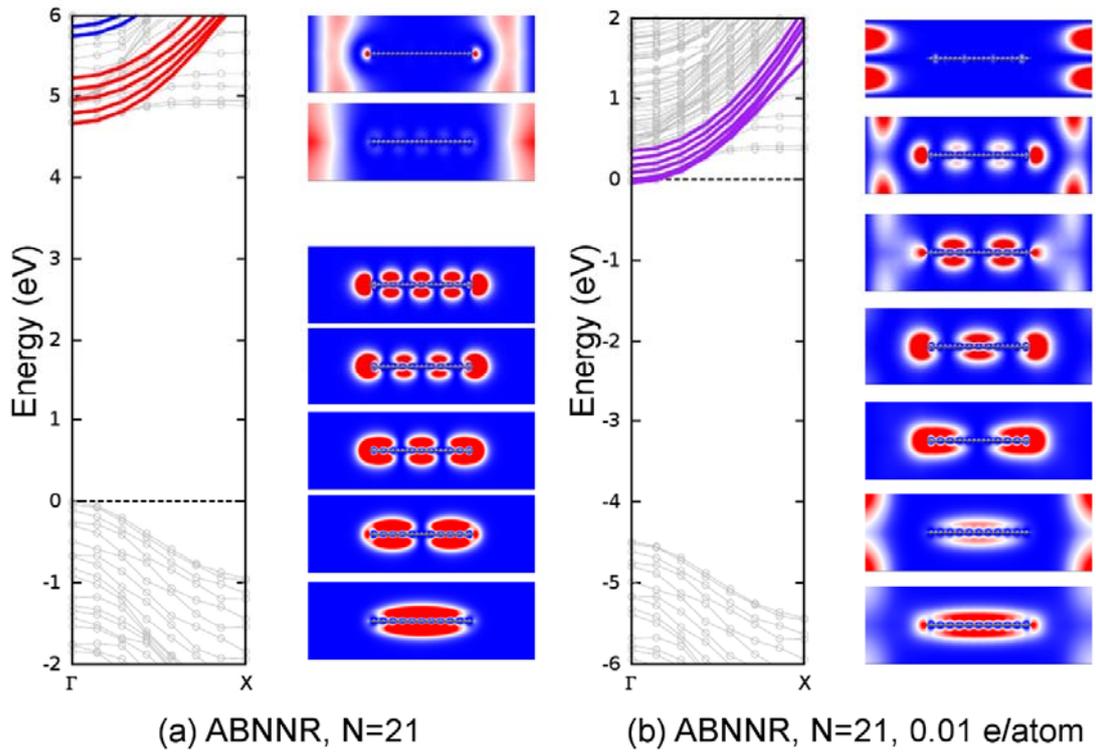

Figure 7

Graphical table of contents

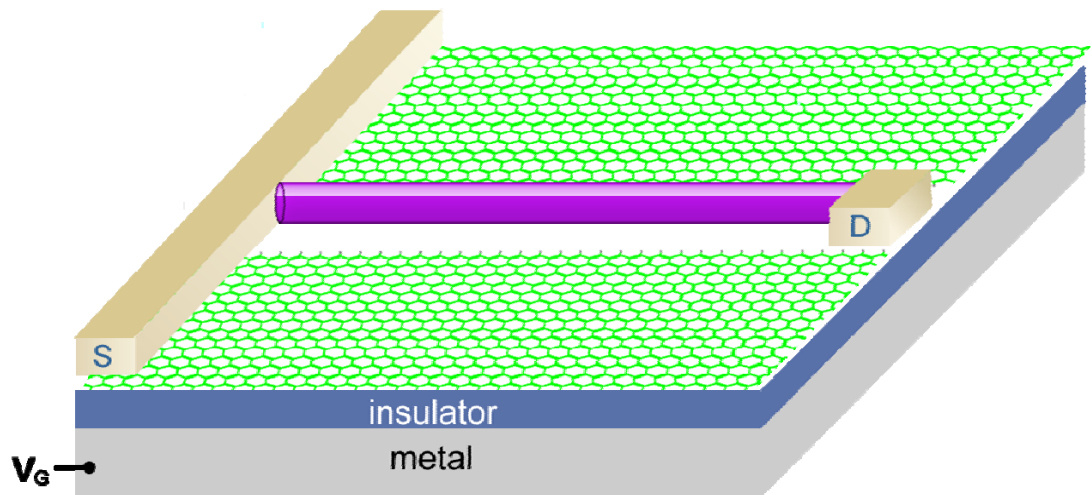

Nearly free electron states in graphene nanoribbon superlattices may be used as tunneling channel upon electron doping.